\documentclass[11pt,a4paper]{article}

\usepackage[utf8]{inputenc}
\usepackage{amsmath,latexsym}
\usepackage{slashed}

\usepackage{tikz}
\usepackage[nomessages]{fp}
\usepackage{jheppub}
\usepackage{glossaries}

\def\beq{\begin{equation}}   
\def\eeq{\end{equation}}
\def\bea{\begin{eqnarray}}  
\def\eea{\end{eqnarray}} 
\def\nn{\nonumber}

\def\f21{{}_2F_{1}}
\def\eps{\epsilon}
\def\order{\mathcal{O}}

\newcommand{\MSb}{$\overline{\mbox{MS}}$}

\def\z#1{{\zeta_{#1}}}

\def\zts{\zeta_3^{\:\!2}}

\def\ca{{C^{}_{\!A}}}
\def\cas{{C^{\,2}_{\!A}}}
\def\cat{{C^{\,3}_{\!A}}}
\def\caf{{C^{\,4}_{\!A}}}
\def\cai{{C^{\,5}_{\!A}}}

\def\cf{{C^{}_F}}
\def\cfs{{C^{\, 2}_F}}
\def\cft{{C^{\, 3}_F}}
\def\cff{{C^{\, 4}_F}}
\def\cfi{{C^{\, 5}_F}}

\def\nf{{n^{}_{\! f}}}

\def\nfs{{n^{\:\!2}_{\! f}}}
\def\nft{{n^{\:\!3}_{\! f}}}
\def\nff{{n^{\:\!4}_{\! f}}}

\def\dfAAna{{\frac{d_A^{\,abcd}d_A^{\,abcd}}{N_A }}} 
\def\dfFAna{{\frac{d_R^{\,abcd}d_A^{\,abcd}}{N_A }}}
\def\dfFFna{{\frac{d_R^{\,abcd}d_R^{\,abcd}}{N_A }}}
\def\dfFAnr{{\frac{d_R^{\,abcd}d_A^{\,abcd}}{N_R }}}
\def\dfFFnr{{\frac{d_R^{\,abcd}d_R^{\,abcd}}{N_R }}}

\def\as(#1){{\alpha_{\rm s}^{\,#1}}}
\def\ar(#1){{a_{\rm s}^{\,#1}}}

\def\Lt(#1){{L_t^{\:\!#1}}}
\def\lntwo(#1){{\ln^{\:\!#1\!}2}}

\def\Gam(#1,#2){{\gamma_{#1}^{\,#2}}}

\def\B(#1,#2){{\beta_{#1}^{\,#2}}}

\def\eps{\epsilon}

\newcommand{\EQN}{\label}

\newcommand{\ed}{\end{document}}

\newcommand{\ice}[1]{\relax}

\newcommand{\re}[1]{(\ref{#1})}

\makeglossaries

\title{Five-loop renormalisation of QCD in covariant gauges}

\author[a]{K. G. Chetyrkin,}
\author[b]{G. Falcioni,}
\author[b]{F. Herzog,}
\author[b]{and J.A.M. Vermaseren}

\affiliation[a]{II Institut für Theoretische Physik,\\ Universität Hamburg, Luruper Chaussee 149, 22761 Hamburg, Germany}
\affiliation[b]{Nikhef Theory Group,\\ Science Park 105, 1098 XG Amsterdam, The Netherlands}

\emailAdd{konstantin.chetyrkin@desy.de}
\emailAdd{gfalcion@nikhef.nl}
\emailAdd{fherzog@nikhef.nl}
\emailAdd{t68@nikhef.nl}

\abstract{
We present the complete set of vertex, wave function and charge renormalisation constants
in QCD in a general simple gauge group and with the complete dependence on the covariant gauge
parameter $\xi$ in the minimal subtraction scheme of conventional dimensional regularisation.
Our results confirm all already known results, which were obtained in the Feynman gauge,
and allow the extraction of other useful gauges such as the Landau gauge. We use these results 
to extract the Landau gauge five-loop anomalous dimensions of the composite operator $A^2$ as well as 
the Landau gauge scheme independent gluon, ghost and fermion propagators at five loops.
}

\begin{document}

\preprint{DESY 17-144, Nikhef 2017-42}

\maketitle

\section{Introduction}

By explaining both the confinement of hadrons as well as the asymptotic freedom of partons observed in 
highly energetic collisions, QCD is one of the phenomenologically richest, most predictive and well tested theories 
of particle and nuclear physics as a whole. Of fundamental importance for making predictions are the anomalous 
dimensions (ADs) and their associated renormalisation constants (RCs). These RCs are indispensable to arrive at finite predictions for observables, such as cross sections and decay rates which can be measured at present and future hadron and lepton collider experiments, where QCD effects play an important role. But the RCs, in particular in Landau gauge, also make up important ingredients when comparing perturbative with non-perturbative results from Lattice determinations.

The computation of QCD RCs in dimensional regularisation \cite{tHooft:1972tcz,Bollini:1972ui} 
have a prominent history, beginning with the Nobel-prize winning discovery of asymptotic freedom in 1973 by Gross, 
Wilczek and Politzer at one loop \cite{Gross:1973id,Politzer:1973fx}. Since then enormous progress has pushed the current status of the art to five loops \cite{Caswell:1974gg, Jones:1974mm, Egorian:1978zx, Tarasov:1980au,Tarrach:1980up,Tarasov:1982gk, Larin:1993tp, vanRitbergen:1997va,Chetyrkin:1997dh,Vermaseren:1997fq, Czakon:2004bu}. At five loops the QCD beta function was first published for the SU(3) gauge group in \cite{Baikov:2016tgj}. This result was confirmed and extended to an arbitrary simple gauge group; the highest power 
in the number of fermions is already known to all orders in perturbation theory \cite{Gracey:1996he}, the subleading power in the number of fermions was presented in \cite{Luthe:2016ima}, while the full result was given in \cite{Herzog:2017ohr}. 

The fermion anomalous mass and field dimensions were similarly first computed for the SU(3) gauge group in \cite{Baikov:2014qja} and then extended to an arbitrary gauge group in \cite{Luthe:2016xec,Baikov:2017ujl}. The Feynman gauge ghost and fermion wave function- and vertex- RCs were presented first for an arbitrary simple gauge group in \cite{Luthe:2017ttc}. While in dimensional regularisation the beta function is entirely gauge independent, the same is not true for the vertex and wave function ADs. The Feynman gauge is often the simplest to work with in practice, nevertheless the Landau gauge often plays a similarly important role, for instance in comparisons of the momentum dependence of propagators and vertex correlators between Lattice simulations and perturbative calculations \cite{Suman:1995zg, Becirevic:1999hj, Becirevic:1999uc, Becirevic:1999sc, vonSmekal:2009ae, Blossier:2010ph, Blossier:2011wk, Bornyakov:2013pha, Blossier:2014kta}. 

An important 1-parameter family of gauges, which includes both the Feynman and the Landau gauge, are the covariant gauges.
Until now not even the linear dependence on this gauge parameter of the RCs was known at five loops\footnote{The linear terms in $\xi$ were published in ref. \cite{Luthe:2017ttg}, which appeared on the same day we submitted this paper. In the concluding section we add a few remarks on this point.}. In this article we address this issue and provide the full gauge parameter dependence in the covariant gauges, for an arbitrary simple gauge group at five loops. This calculation was possible by combining two extremely powerful approaches. Using the global $R^*$-method \cite{Caswell:1981ek,Chetyrkin:1982nn,Chetyrkin:1984xa,Smirnov:1986me,Chetyrkin:2017ppe}, see also \cite{Steinhauser:2002rq} for a review of this technique, the required five-loop self energy and vertex integrals are  written in terms of tadpole integrals, whose lines are, apart from one single massive line, all massless. 
These tadpoles integrals, which can always be factorised into products of four-loop massless propagator integrals 
times trivial one-loop tadpole integrals, are subsequently evaluated with the {\sc Forcer} program \cite{Ruijl:2017cxj,Ueda:2016yjm,Ueda:2016sxw}.
This approach, which is at least an order of magnitude faster than any of the alternative approaches used so far, 
such as the local $R^*$-method recently developed in \cite{Herzog:2017bjx}, enabled us to compute the full gauge parameter 
dependence of the ghost and fermion self energy and ghost-gluon vertex correlator. 
This has allowed us to reconstruct the full gauge parameter dependence of all QCD RCs at five loops and constitutes the main new result of this work. These results were previously known only at the four-loop level \cite{Chetyrkin:2004mf, Luthe:2017ttc}.
For the special case of the Landau gauge to all orders in perturbation theory the gluon and ghost ADs are known at leading power in the number of fermions \cite{Gracey:1993ua} while the quark ADs are known to subleading power \cite{Gracey:1993ua,Ciuchini:1999wy}; these results provide important cross checks for the parts of the results presented here.

The application of the global $R^*$-method to this problem was conceptually more challenging than in the case of the SU(3) 
gauge group \cite{Baikov:2016tgj}. We were able to solve these conceptual problems and in addition refined and simplified the global R* method by making use of the QCD Ward Identities. We will present the details of this method in a future publication \cite{tobepublished}.

This paper is organised as follows. In Section \ref{Setup} we will set up our notations and conventions. 
Explicit results will be given for the case of the Landau gauge in section \ref{Results}. General gauge results can be found in an ancillary file. We apply these results to obtain the scheme independent gluon, ghost and fermion propagators as well as the anomalous dimension (AD) of the composite operator $A^2$ in section \ref{sec:Applications}. 
We conclude in section \ref{Conclusions}.

\section{Setup}
\label{Setup}
In the covariant gauge the QCD Lagrangian with $n_f$  quark flavors reads:
\beq
{\cal L}  =  
\sum_{f=1}^{n_f} \bar \psi^f ( \mathrm{i}  \slash{\hspace*{-2.34mm} D}  -m_f) \psi^f
-\frac{1}{4} G^a_{\mu\nu} G_{a\mu\nu}
\EQN{lagr:r}
+ \partial^\mu \bar c _a D_\mu^{ab} c_b 
- \frac{1}{2\xi_L} (\partial^\mu A^a_\mu)^2
{}\,.
\eeq
Here the gluon field strength is defined by
\beq
G^a_{\mu\nu}  =  \partial_\mu A^a_\nu - \partial_\nu A^a_\mu +g\,(A_\mu \times A_\nu)^a,
\qquad (A \times B)^a = f^{abc} A^b B^c,
\eeq
the covariant derivatives are given by
\beq
D_\mu^{ab} = \delta^{ab} \partial_\mu - g f^{abc}  A^c_\mu\,,
\qquad D_{\mu} = \partial_\mu - \mathrm{i} g A^a_\mu T_R^a\,.
\eeq
and are using the standard Feynman slash notation, e.g. $\slashed{p}=p_\mu \gamma^\mu$.
The quark field $\psi^f_i$, with mass $m_f$, transforms in an arbitrary representation $R$ of an arbitrary simple gauge group $G$.
The gauge fields $A^a_\mu$ will be taken in the adjoint representation 
of the gauge group $G$. $(T_R)^a_{ij}$ are the generators in the representation of the fermions and $f^{abc}$ are the 
structure constants of the corresponding Lie algebra. Here $c^a$ denote the
ghost fields and $ \xi_L $ acts as the gauge parameter such that $ \xi_L =0$ corresponds
to the Landau gauge, while $\xi_L=1$ corresponds to the Feynman gauge.

Supplementing eq.\re{lagr:r} with all counterterms necessary to remove possible UV divergences 
from Green functions, one arrives at the {\em bare}  
QCD Lagrangian, written in terms of the  renormalised quantities\footnote{
For simplicity we set  the t${}'$ Hooft mass $\mu=1$ in  eq. (\ref{lagr:2})
below.
} :
\begin{eqnarray}
{\cal L}_0 & = &
Z_2\sum_{f=1}^{n_f} 
\bar \psi^f ( \mathrm{i}\,\,  \slash{ \hspace*{-2.34mm} \partial } 
+ g Z^{\psi\psi g}_1 Z_2^{-1}\slash{ \hspace*{-2.67mm} A} - Z_m m_f )\, \psi^f
+ g\, Z_1^{ccg} \, \partial^\mu \bar c \, (A \times   c  )
\nonumber
\\
& &
- \frac{1}{4} g^2\, Z^{4g}_1\, ( A_\mu \times A_\nu)^2
- Z_\xi^{2g}\, \frac{1}{2  \xi_L }  ( \partial _\nu A_\mu)^2
+ Z^c_3\, \partial_\nu \bar c  \, (\partial_\nu c )
\EQN{lagr:2}
\\
&&
- \frac{1}{4} Z_3\, ( \partial _{\mu}A_{\nu} -  \partial _{\nu}A_{\mu})^2
- \frac{1}{2}g\, Z_1^{3g} \, ( \partial _{\mu}A^a_{\nu} -  \partial _{\nu}A^a_{\mu})
\,  ( A_{\mu} \times A_{\nu})^a 
\nonumber
{}\,.
\end{eqnarray}
The RC $Z_\xi^{2g}$ is expressed  through the RCs of the gauge fixing parameter
$ \xi_L $ as follows
\[
\xi_{L,0} =Z_\xi  \xi_L , \ \ \ Z_\xi^{2g} = Z_3/Z_\xi
{}.
\]
The wave-function RCs, $Z_3, Z_2$ and $Z_3^{c}$,
arise in the relations between renormalised and bare gluon, quark and ghost fields, as follows
\begin{equation} 
A^{a \mu}_0 = \sqrt{Z_3}\ A^{a \mu},
\qquad
\psi^f_0  = \sqrt{Z_2}\ \psi^f,
\qquad
c^{a}_0   = \sqrt{Z_3^{c}}\ c^{a} 
{}.
\end{equation} 
Similarly $Z_m$ and $Z_g$ arise in the relations between bare and renormalised charge 
and fermion mass parameters:
\beq
g_0=\mu^{\eps} Z_g\, g\,, \qquad m_0=Z_m\,m\,.
\eeq
The full set of vertex RCs
\begin{equation} 
\label{eq:ZVset}
Z^V_1, \ \ \ V\in \{\mathrm{3g,\ 4g,\  c  c  g  , \ \psi  \psi g}\}
\end{equation} 
serve to renormalise the three-gluon, four-gluon, ghost-ghost-gluon and quark-quark-gluon vertex functions respectively. 
Gauge invariance leads to the Slavnov-Taylor identities among the RCs:
\beq
Z_\xi = Z_3,\qquad Z_g = \frac{(Z_1^{4g})^{1/2}}{ Z_3} = \frac{Z_1^{3g}} {Z_3^{3/2}} = \frac{Z_1^{ccg}} {(Z_3)^{1/2} Z_3^c} = \frac{Z_1^{\psi\psi g}}{ (Z_3)^{1/2} Z_2}\,.
\EQN{WI:qqg}
\eeq
To arrive at the complete set of RCs it therefore suffices
to determine all RCs from the knowledge of only $Z_m,Z_2,Z_3,Z_{3}^c$ and $Z_1^{ccg}$. 
In the minimal subtraction scheme (MS), the RCs $Z_i$ are pure poles in the
dimensional regularization parameter $\epsilon$
\begin{equation}
  Z_i=1+\sum_{k=1}^\infty\frac{z_i^{(k)}}{\epsilon^k},
  \EQN{eq:Laurent}
\end{equation}
and the coefficients $z_i^{(k)}$ do not depend explicitly on mass
scales \cite{Collins:1974bg}. These features greatly simplify the
calculation of the RCs, by imposing the finiteness condition on the
propagators and vertex functions computed with the Lagrangian of
eq. (\ref{lagr:2}), that is
\begin{align}
 K_\epsilon\bigg[Z_3\left(1+\Pi_B(g_0,q^2)\right)\bigg]&=0,\\
 K_\epsilon\bigg[Z_{3}^{c}\left(1+\tilde{\Pi}_B(g_0,q^2)\right)
   \bigg]&=0,\\
 K_\epsilon\bigg[Z_2\left(1+\Sigma_B(g_0,q^2)\right)\bigg]&=0,\\
 K_\epsilon\bigg[Z_1^{ccg}\left(1+\delta\Gamma_B^{ccg}(g_0,q^2)
   \right)\bigg]&=0,
 \EQN{eq:ren}
\end{align}
where the operator $K_\epsilon$ extracts the pole part of the generic
Laurent expansion
\begin{equation}
  K_\epsilon\bigg[\sum_{k=-\infty}^{\infty} c_k\,\epsilon^k\bigg]
  = \sum_{k=-\infty}^{-1}c_k\epsilon^k,
\end{equation}
while $\Pi_B$, $\tilde{\Pi}_B$, $\Sigma_B$ denote respectively the
bare 1PI self energies of the gluon, of the ghost and of the fermion
and $\delta\Gamma_B^{ccg}$ is the bare ghost-gluon vertex correction.
The constant $Z_m$ is derived in a similar way from the finiteness of
the renormalized operator $\overline{\Psi}\Psi$
\begin{align}
  Z_m&=\frac{Z_{\overline{\Psi}\Psi}}{Z_2},\quad
  K_\epsilon\bigg[Z_{\overline{\Psi}\Psi}
    \left(1+\delta\Gamma_B^{\overline{\Psi}\Psi}(g_0)\right)\bigg]=0,
\end{align}
where $\delta\Gamma_B^{\overline{\Psi}\Psi}(g_0)$ is the loop
correction of the vertex of the operator $\overline{\Psi}\Psi$.
Finally, the RCs give immediately the ADs that govern
the renormalisation group equation of the theory: by using the
independence of the RCs on mass scales, the
evolution equations become
\begin{align}
\begin{split}
  \EQN{eq:gamma}
  \frac{d}{d\log(\mu^2)}Z_i&=\bigg[
  \left(-\epsilon+\beta\right)a\frac{\partial}{\partial a}
  +\gamma_3\,\xi_L\frac{\partial}{\partial\xi_L}\bigg]Z_i
  = -\gamma_i\, Z_i,
\end{split}
\end{align}
where $\beta$ and $\gamma_3$ are the ADs related to
the renormalised coupling constant 
\beq
a(\mu)\equiv\left(\frac{g(\mu)}{4\pi}\right)^2=\frac{\alpha_s(\mu)}{4\pi} 
\eeq
and to the gluon renormalization, respectively
\begin{equation}
  \beta=-\frac{d\log((Z_g)^2)}{d\log(\mu^2)} 
  \equiv-\frac{d\log(Z_a)}{d\log(\mu^2)}=\frac{d\log(a)}{d\log(\mu^2)}+\eps,\qquad
  \gamma_3=-\frac{d\log(Z_3)}{d\log(\mu^2)}.
\end{equation}
The expansion of eq. (\ref{eq:gamma}) in powers of $\epsilon$ by means of eq. (\ref{eq:Laurent})
gives the relation between the poles of the RCs and ADs of the theory
\begin{align}
  \begin{split}
    \gamma_i&=a\frac{\partial}{\partial a}z_i^{(1)},\qquad
    a\frac{\partial z_i^{(k+1)}}{\partial a}=
    \bigg[\beta\,a\frac{\partial}{\partial a}+
      \gamma_3\,\xi_L\frac{\partial}{\partial\xi_L}+\gamma_i\bigg]
    z_i^{(k)}.
  \EQN{eq:zpoles}  
  \end{split}
\end{align}

\section{Calculation and results}
\label{Results}

For the computation of the RCs we employ the global $R^*$-method. This method, which 
has already been used in many important calculations in multi-loop QCD, see e.g. \cite{Chetyrkin:2004mf, Baikov:2005rw, Baikov:2012zm, Baikov:2012er, Baikov:2014qja}, was also used in the calculation of the 5-loop QCD RCs for the SU(3) gauge group in \cite{Baikov:2016tgj}.
Generally the global $R^*$ method relies on a global infrared rearrangement of 
a correlator by inserting a mass $M$ into one or two propagators attached to one of the external vertices.
Subsequently the correlator is expanded around the limit $M\to\infty$, or equivalently the limit 
of vanishing external momenta. In this way the 5-loop correlator becomes a ``one-mass-tadpole'', 
which can be evaluated as a product of a 4-loop massless propagator integral times a trivial 
1-loop tadpole. This makes it possible to use the FORCER program \cite{Ruijl:2017cxj} in order to efficiently 
perform the reduction to Master integrals.

To extract the correct RCs from ``one-mass-tadpole'' correlators does however require a subtle 
derivation of global renormalisation counterterms. While the application of the global $R^*$ method is comparably 
straightforward for fermion or ghost vertices and self energies, extra complications arise for the gluon propagator.
These complications emerge from the fact that the type of the external gluon vertex can be any of 
the four QCD vertices in eq.(\ref{eq:ZVset}), rather than just a single type in the case of external 
fermion or ghost vertices. This leads in particular to the phenomenon of operator mixing under renormalisation 
and further demands the introduction of new (renormalisable) operators into the QCD Lagrangian.

To extend the global $R^*$-method to a general gauge group required the introduction 
of yet new 4-gluon operators (whose color structures could not appear in the SU(3) case) for the construction 
of the global renormalisation counterterms. To obtain the results presented in this article, we have both extended 
and refined the global $R^*$ method in order to deal with these extra complications. 
Since the details of the method are rather subtle, we will present them in a future publication \cite{tobepublished}.

All required Feynman diagrams were generated with QGRAF \cite{Nogueira:1991ex}. Further symbolic manipulations where performed with  the computer algebra system {\sc Form} \cite{Vermaseren:2000nd,Tentyukov:2007mu,Kuipers:2012rf,Ruijl:2017dtg}. The reduction to Master integrals is performed using the {\sc Forcer} program \cite{Ruijl:2017cxj,Ueda:2016yjm,Ueda:2016sxw}. Our computation of colour factors relied on the {\sc Form} package introduced in \cite{vanRitbergen:1998pn}. The total computation of all required components took approximately two months on a single computer with $32$ cores.

Using the methods described above we have been able to directly compute $Z_{3}^c,Z_{1}^{\mathrm{ccg}}$ and $Z_2$ 
to all powers -- and $Z_3$ to leading power -- in the gauge-parameter up to five loops. Use of the Slavnov-Taylor identities 
then allows us to reconstruct the entire gauge parameter dependence of $Z_3$, using the explicitly computed linear dependence as 
a cross check. 

In the following we present analytic results for the 5-loop coefficients of the ADs 
$\gamma_2$,$\gamma_3$ and $\gamma_{3c}$, each of which have perturbative expansions of the form,
\beq
\gamma_i=-\sum_{k=0}^\infty a^{k+1} (\gamma_i)_k \,,
\eeq
in the Landau gauge ($\xi_L=0$):
\bea
(\gamma_2)_0&&= 0 ,\qquad
(\gamma_2)_1 =
       - 2\, { \cf\, \* \nf\, \* T_R} 
       - { 3 \over 2 }\, {\cfs }
        + { 25 \over 4 }\,{ \ca\, \* \cf }\,,
\eea
\bea
(\gamma_2)_2 &&=
        { 20 \over 9 }\,{ \cf\, \* \nfs\, \* T_R^2 }
        + 3\, \cfs\, \* \nf\, \* T_R 
        + { 3 \over 2 }\, \cft 
        - { 287 \over 9 }\, {\ca\, \* \cf\, \* \nf\, \* T_R }\nn\\
&&
       + \ca\, \* \cfs \, \*  \Big[
          - { 143 \over 4 }
          + 12\, \* \z3
          \Big]
       + \cas\, \* \cf \, \*  \Big[
           { 9155 \over 144 }
          - { 69 \over 8 }\, \* \z3
          \Big]\,,
\eea
\bea
(\gamma_2)_3 &&=
       { \dfFAnr} \, \*  \Big[
          - 67
          - { 855 \over 8 }\, \* \z5
          + { 1113 \over 8 }\, \* \z3
          \Big]
       + 128\,\nf \* \dfFFnr \nn\\
&&
       +  { 280 \over 81 }\,\cf\, \* \nft\, \* T_R^3 
       + \cfs\, \* \nfs\, \* T_R^2 \, \*  \Big[
           { 304 \over 9 }
          - 32\, \* \z3
          \Big]
       + \cft\, \* \nf\, \* T_R \, \*  \Big[
           { 76 \over 3 }
          - 64\, \* \z3
          \Big]\nn\\
&&
       + \cff \, \*  \Big[
          - { 1027 \over 8 }
          + 640\, \* \z5
          - 400\, \* \z3
          \Big]
       + \ca\, \* \cf\, \* \nfs\, \* T_R^2 \, \*  \Big[
           { 293 \over 9 }
          + 32\, \* \z3
          \Big]\nn\\
&&
       + \ca\, \* \cfs\, \* \nf\, \* T_R \, \*  \Big[
          - { 53 \over 18 }
          + 24\, \* \z4
          + 160\, \* \z5
          \Big]
       + \ca\, \* \cft \, \*  \Big[
           { 5131 \over 12 }
          - 1440\, \* \z5
          + 848\, \* \z3
          \Big]\nn\\
&&
       + \cas\, \* \cf\, \* \nf\, \* T_R \, \*  \Big[
          - { 18371 \over 54 }
          - { 69 \over 4 }\, \* \z4
          - 80\, \* \z5
          - { 3 \over 2 }\, \* \z3
          \Big]
       + \cas\, \* \cfs \, \*  \Big[
          - { 23885 \over 36 }
          - 66\, \* \z4\nn\\
&&
          + 785\, \* \z5
          - { 421 \over 2 }\, \* \z3
          \Big]
       + \cat\, \* \cf \, \*  \Big[
           { 95261 \over 162 }
          + { 759 \over 16 }\, \* \z4
          - { 4145 \over 32 }\, \* \z5
          - { 6209 \over 64 }\, \* \z3
          \Big]\,,
\eea
\bea
(\gamma_2)_4&& =
        \cf\, \* \dfAAna \, \*  \Big[
          - { 1985 \over 24 }
          + { 781753 \over 192 }\, \* \z7
          - { 1458845 \over 384 }\, \* \z5
          + { 135731 \over 192 }\, \* \z3
          + { 3577 \over 64 }\, \* \zts
          \Big]\nn\\
&&
       + T_R\, \* \nf\, \* \dfFAnr \, \*  \Big[
        { 6200 \over 9 }
          - { 1425 \over 4 }\, \* \z6
          + { 27377 \over 6 }\, \* \z7
          + { 1113 \over 4 }\, \* \z4
          - { 9915 \over 2 }\, \* \z5\nn\\
&&
          - { 2468 \over 3 }\, \* \z3
          + { 91 \over 2 }\, \* \zts
          \Big]
       + T_R\, \* \nfs\, \* \dfFFnr \, \*  \Big[
       - { 7360 \over 9 }
          + 640\, \* \z5
          + { 704 \over 3 }\, \* \z3
          \Big]\nn\\
&&
       + \cf\, \* \nff\, \* T_R^4 \, \*  \Big[
           { 1328 \over 243 }
          - { 256 \over 27 }\, \* \z3
          \Big]
       + \cf\, \* \dfFAnr \, \*  \Big[
           { 113 \over 6 }
          - { 125447 \over 8 }\, \* \z7\nn\\
&&
          + 1015\, \* \z5
          + 17554\, \* \z3
          - 4884\, \* \zts
          \Big]
          + \cf\, \* \nf\, \* \dfFFnr \, \*  \Big[
          - { 5984 \over 3 }
          - 8680\, \* \z7\nn
\eea
\bea          
\phantom{(\gamma_2)_4}&&
          + 18080\, \* \z5
          - 12096\, \* \z3
          + 3648\, \* \zts
          \Big]
       + \cfs\, \* \nft\, \* T_R^3 \, \*  \Big[
          - { 2636 \over 243 }
          - 64\, \* \z4
          + { 832 \over 9 }\, \* \z3
          \Big]\nn\\
&&
       + \cft\, \* \nfs\, \* T_R^2 \, \*  \Big[
          - { 2497 \over 27 }
          - 128\, \* \z4
          + 320\, \* \z5
          + { 400 \over 9 }\, \* \z3
          \Big]
       + \cff\, \* \nf\, \* T_R \, \*  \Big[
           { 29209 \over 36 }\nn\\
&&
          + { 6400 \over 3 }\, \* \z6
          - 800\, \* \z4
          - { 46880 \over 9 }\, \* \z5
          + { 22496 \over 9 }\, \* \z3
          + { 1024 \over 3 }\, \* \zts
          \Big]
       + \cfi \, \*  \Big[
           { 4977 \over 8 }\nn\\
&&
          - 47628\, \* \z7
          + 22600\, \* \z5
          + 16000\, \* \z3
          + 2496\, \* \zts
          \Big]
       + \ca\, \* \dfFAnr \, \*  \Big[
          - { 173959 \over 144 }\nn\\
&&
          + { 15675 \over 16 }\, \* \z6
          + { 3016307 \over 256 }\, \* \z7
          - { 12243 \over 16 }\, \* \z4
          + { 609425 \over 96 }\, \* \z5
          - { 574393 \over 32 }\, \* \z3
          + { 16935 \over 4 }\, \* \zts
          \Big]\nn\\
&&
       + \ca\, \* \nf\, \dfFFnr \, \*  \Big[
        { 33464 \over 9 }
          + { 23632 \over 3 }\, \* \z7
          - { 48640 \over 3 }\, \* \z5
          + 8992\, \* \z3
          - 2320\, \* \zts
          \Big]\nn\\
&&
       + \ca\, \* \cf\, \* \nft\, \* T_R^3 \, \*  \Big[
          - { 3566 \over 243 }
          + 64\, \* \z4
          - { 1984 \over 27 }\, \* \z3
          \Big]
       + \ca\, \* \cfs\, \* \nfs\, \* T_R^2 \, \*  \Big[
           { 101485 \over 162 }\nn\\
&&
          + { 1600 \over 3 }\, \* \z6
          + 176\, \* \z4
          - { 3712 \over 3 }\, \* \z5
          - { 6160 \over 9 }\, \* \z3
          + { 256 \over 3 }\, \* \zts
          \Big]
       + \ca\, \* \cft\, \* \nf\, \* T_R \, \*  \Big[
          - { 167263 \over 108 }\nn\\
&&
          - 4800\, \* \z6
          - 13944\, \* \z7
          + 2120\, \* \z4
          + { 58720 \over 3 }\, \* \z5
          - { 25804 \over 9 }\, \* \z3
          - 64\, \* \zts
          \Big]\nn\\
&&
      + \ca\, \* \cff \, \*  \Big[
          - { 835739 \over 144 }
          - { 17600 \over 3 }\, \* \z6
          + 123977\, \* \z7
          + 2200\, \* \z4
          - { 248960 \over 9 }\, \* \z5
          - { 530884 \over 9 }\, \* \z3\nn\\
&&
          - { 24632 \over 3 }\, \* \zts
          \Big]
       + \cas\, \* \cf\, \* \nfs\, \* T_R^2 \, \*  \Big[
           { 120037 \over 162 }
          - { 800 \over 3 }\, \* \z6
          - { 441 \over 2 }\, \* \z7
          - 179\, \* \z4
          + { 3584 \over 9 }\, \* \z5\nn\\
&&
          + { 3140 \over 3 }\, \* \z3
          - { 128 \over 3 }\, \* \zts
          \Big]
       + \cas\, \* \cfs\, \* \nf\, \* T_R \, \*  \Big[
           { 717409 \over 432 }
          + 1150\, \* \z6
          + { 42203 \over 3 }\, \* \z7
          - { 1411 \over 4 }\, \* \z4\nn\\
&&
          - { 95792 \over 9 }\, \* \z5
          - { 14287 \over 24 }\, \* \z3
          - 1214\, \* \zts
          \Big]
       + \cas\, \* \cft \, \*  \Big[
           { 827215 \over 72 }
          + 13200\, \* \z6
          - { 1789067 \over 16 }\, \* \z7\nn\\
&&
          - 4664\, \* \z4
          - { 188795 \over 12 }\, \* \z5
          + { 1365227 \over 18 }\, \* \z3
          + { 18097 \over 2 }\, \* \zts
          \Big]
       + \cat\, \* \cf\, \* \nf\, \* T_R \, \*  \Big[
          - { 31919039 \over 7776 }\nn\\
&&
          + { 4825 \over 16 }\, \* \z6
          - { 440419 \over 144 }\, \* \z7
          - { 8705 \over 32 }\, \* \z4
          + { 28721 \over 18 }\, \* \z5
          - { 144377 \over 864 }\, \* \z3
          + { 4067 \over 6 }\, \* \zts
          \Big]\nn\\
&&
       + \cat\, \* \cfs \, \*  \Big[
          - { 42214139 \over 3888 }
          - { 43175 \over 6 }\, \* \z6
          + { 9074513 \over 192 }\, \* \z7
          + { 3815 \over 4 }\, \* \z4
          + { 5957573 \over 288 }\, \* \z5\nn\\
&&
          - { 5503507 \over 144 }\, \* \z3
          - { 78041 \over 24 }\, \* \zts
          \Big]
       + \caf\, \* \cf \, \*  \Big[
           { 368712343 \over 62208 }
          + { 227975 \over 192 }\, \* \z6\nn\\
&&
          - { 312820991 \over 36864 }\, \* \z7
          + { 87067 \over 128 }\, \* \z4
          - { 16237513 \over 3072 }\, \* \z5
          + { 46196783 \over 6912 }\, \* \z3
          + { 23555 \over 128 }\, \* \zts
          \Big]\,,
\eea

\bea
(\gamma_3)_0&&=          
        { 4 \over 3 }\,\nf\, \* T_R 
       - { 13 \over 6 }\, \ca \,,
\eea 
\bea
(\gamma_3)_1 &&=               
         4 \,\cf\, \* \nf\, \* T_R 
       + 5\, \ca\, \* \nf\, \* T_R 
       - { 59 \over 8 } \cas \,,
\eea
\bea
(\gamma_3)_2 &&=
       - { 44 \over 9 }\, \cf\, \* \nfs\, \* T_R^2 
       - 2\, \cfs\, \* \nf\, \* T_R 
       - { 76 \over 9 }\, \ca\, \* \nfs\, \* T_R^2 
       + \ca\, \* \cf\, \* \nf\, \* T_R \, \*  \Big[
           { 5 \over 18 }\nn\\
&&          
          + 24\, \* \z3
          \Big]
       + \cas\, \* \nf\, \* T_R \, \*  \Big[
           { 911 \over 18 }
          - 18\, \* \z3
          \Big]
       + \cat \, \*  \Big[
          - { 9965 \over 288 }
          + { 9 \over 16 }\, \* \z3
          \Big]\,,
\eea
\bea
(\gamma_3)_3 &&=
        \dfAAna \, \*  \Big[
           { 659 \over 144 }
          - { 10185 \over 64 }\, \* \z5
          - { 989 \over 12 }\, \* \z3
          \Big]
       + \nf\, \* \dfFAna \, \*  \Big[
          - { 512 \over 9 }
          + 120\, \* \z5\nn\\
&&          
          + { 1376 \over 3 }\, \* \z3
          \Big]
       + \nfs\, \* \dfFFna \, \*  \Big[
           { 704 \over 9 }
          - { 512 \over 3 }\, \* \z3
          \Big]
       + \cf\, \* \nft\, \* T_R^3 \, \*  \Big[
          - { 1232 \over 243 }
          \Big]\nn\\
&&          
       + \cfs\, \* \nfs\, \* T_R^2 \, \*  \Big[
          - { 1352 \over 27 }
          + { 704 \over 9 }\, \* \z3
          \Big]
       + \cft\, \* \nf\, \* T_R \, \*  \Big[
          - 46
          \Big]
       + \ca\, \* \nft\, \* T_R^3 \, \*  \Big[
          - { 1420 \over 243 }
          + { 64 \over 9 }\, \* \z3
          \Big]\nn\\
&&          
       + \ca\, \* \cf\, \* \nfs\, \* T_R^2 \, \*  \Big[
          - { 15082 \over 243 }
          + 48\, \* \z4
          - { 1168 \over 9 }\, \* \z3
          \Big]
       + \ca\, \* \cfs\, \* \nf\, \* T_R \, \*  \Big[
           { 10847 \over 54 }
          - 240\, \* \z5\nn\\
&&          
          + { 980 \over 9 }\, \* \z3
          \Big]
       + \cas\, \* \nfs\, \* T_R^2 \, \*  \Big[
          - { 6674 \over 81 }
          - 36\, \* \z4
          + { 316 \over 9 }\, \* \z3
          \Big]
       + \cas\, \* \cf\, \* \nf\, \* T_R \, \*  \Big[
          - { 146831 \over 972 }\nn\\
&&          
          - 132\, \* \z4
          + 120\, \* \z5
          + { 2240 \over 9 }\, \* \z3
          \Big]
       + \cat\, \* \nf\, \* T_R \, \*  \Big[
           { 318907 \over 864 }
          + { 801 \over 8 }\, \* \z4
          + 110\, \* \z5
          - { 3563 \over 12 }\, \* \z3
          \Big]\nn\\
&&          
       + \caf \, \*  \Big[
          - { 10655437 \over 62208 }
          - { 99 \over 32 }\, \* \z4
          - { 47665 \over 512 }\, \* \z5
          + { 50669 \over 768 }\, \* \z3
          \Big]\,,
\eea
\bea
(\gamma_3)_4&&=      
        \nf\, \* T_R\, \* \dfAAna \, \*  \Big[
          - { 3043 \over 36 }
          - { 16975 \over 32 }\, \* \z6
          + { 5285 \over 4 }\, \* \z7
          - { 95 \over 2 }\, \* \z4
          - { 232535 \over 288 }\, \* \z5
          + { 505937 \over 288 }\, \* \z3\nn\\
&&          
          - { 2017 \over 16 }\, \* \zts
          \Big]
       + \nfs\, \* T_R\, \* \dfFAna \, \*  \Big[
           { 3680 \over 9 }
          + 400\, \* \z6
          + 640\, \* \z4
          - { 4720 \over 9 }\, \* \z5
          - { 37664 \over 9 }\, \* \z3\nn\\
&&          
          + 96\, \* \zts
          \Big]
       + \nft\, \* T_R\, \* \dfFFna \, \*  \Big[
          - { 3520 \over 9 }
          - 256\, \* \z4
          - { 1280 \over 3 }\, \* \z5
          + { 2624 \over 3 }\, \* \z3
          \Big]\nn\\
&&          
       + \cf\, \* \nf\, \* \dfFAna \, \*  \Big[
           { 944 \over 3 }
          - 476\, \* \z7
          - { 3580 \over 3 }\, \* \z5
          - { 1544 \over 3 }\, \* \z3
          - 288\, \* \zts
          \Big]\nn\\
&&          
       + \cf\, \* \nfs\, \* \dfFFna \, \*  \Big[
          - { 4160 \over 3 }
          + { 12800 \over 3 }\, \* \z5
          - { 5120 \over 3 }\, \* \z3
          \Big]
       + \cf\, \* \nff\, \* T_R^4 \, \*  \Big[
           { 856 \over 243 }
          + { 128 \over 27 }\, \* \z3
          \Big]       \nn\\
&&          
       + \cfs\, \* \nft\, \* T_R^3 \, \*  \Big[
           { 9922 \over 81 }
          + { 352 \over 3 }\, \* \z4
          - { 7616 \over 27 }\, \* \z3
          \Big]
       + \cft\, \* \nfs\, \* T_R^2 \, \*  \Big[
           { 5018 \over 9 }
          - { 4640 \over 3 }\, \* \z5
          + { 2144 \over 3 }\, \* \z3
          \Big]\nn\\
&&
       + \cff\, \* \nf\, \* T_R \, \*  \Big[
           { 4157 \over 6 }
          + 128\, \* \z3
          \Big]
       + \ca\, \* \dfAAna \, \*  \Big[
           { 204521 \over 1152 }
          + { 186725 \over 128 }\, \* \z6\nn\\
&&          
          + { 6235831 \over 4096 }\, \* \z7
          + { 2101 \over 8 }\, \* \z4
          + { 140675 \over 256 }\, \* \z5
          - { 5513387 \over 1536 }\, \* \z3
          - { 16887 \over 512 }\, \* \zts
          \Big]\nn\\
&&          
       + \ca\, \* \nf\, \* \dfFAna \, \*  \Big[
          - { 9752 \over 9 }
          - 1100\, \* \z6
          - 784\, \* \z7
          - 2072\, \* \z4
          - { 32885 \over 9 }\, \* \z5
          + { 93551 \over 9 }\, \* \z3\nn\\
&&          
          + 893\, \* \zts
          \Big]
       + \ca\, \* \nfs\, \* \dfFFna \, \*  \Big[
           { 6328 \over 3 }
          + 800\, \* \z4
          - { 6560 \over 9 }\, \* \z5
          - { 37720 \over 9 }\, \* \z3
          + 256\, \* \zts
          \Big]\nn\\
&&          
       + \ca\, \* \nff\, \* T_R^4 \, \*  \Big[
          - { 2476 \over 243 }
          + { 128 \over 9 }\, \* \z4
          \Big]
       + \ca\, \* \cf\, \* \nft\, \* T_R^3 \, \*  \Big[
           { 16973 \over 243 }
          - { 704 \over 3 }\, \* \z4
          + { 256 \over 3 }\, \* \z5\nn\\
&&          
          + { 7072 \over 27 }\, \* \z3
          \Big]
       + \ca\, \* \cfs\, \* \nfs\, \* T_R^2 \, \*  \Big[
          - { 135571 \over 108 }
          - 800\, \* \z6
          + { 44 \over 3 }\, \* \z4
          + 2800\, \* \z5
          - { 16900 \over 27 }\, \* \z3\nn\\
&&          
          + 448\, \* \zts
          \Big]
       + \ca\, \* \cft\, \* \nf\, \* T_R \, \*  \Big[
          - 3759
          + 3360\, \* \z7
          - { 1340 \over 3 }\, \* \z5
          - { 6164 \over 3 }\, \* \z3
          \Big]\nn
          \\
&&          
       + \cas\, \* \nft\, \* T_R^3 \, \*  \Big[
           { 61223 \over 486 }
          + { 392 \over 9 }\, \* \z4
          - { 2176 \over 9 }\, \* \z5
          + { 5792 \over 27 }\, \* \z3
          \Big]
       + \cas\, \* \cf\, \* \nfs\, \* T_R^2 \, \*  \Big[
           { 92417 \over 486 }\nn\\
&&          
          + 400\, \* \z6
          + { 3800 \over 3 }\, \* \z4
          - { 5008 \over 3 }\, \* \z5
          - { 20596 \over 27 }\, \* \z3
          - 928\, \* \zts
          \Big]
       + \cas\, \* \cfs\, \* \nf\, \* T_R \, \*  \Big[
           { 1088107 \over 162 }\nn\\
&&          
          + 2200\, \* \z6
          - 3360\, \* \z7
          - { 1892 \over 3 }\, \* \z4
          - 5650\, \* \z5
          + { 183419 \over 54 }\, \* \z3
          + 880\, \* \zts
          \Big]\nn\eea
\bea          
\phantom{(\gamma_2)_4}&&
       + \cat\, \* \nfs\, \* T_R^2 \, \*  \Big[
          - { 575909 \over 324 }
          + { 1100 \over 3 }\, \* \z6
          - { 1929 \over 2 }\, \* \z4
          + { 2881 \over 4 }\, \* \z5
          + { 39451 \over 216 }\, \* \z3
          + { 1328 \over 3 }\, \* \zts
          \Big]\nn\\
&&          
       + \cat\, \* \cf\, \* \nf\, \* T_R \, \*  \Big[
          - { 352669 \over 72 }
          - 1100\, \* \z6
          + { 16681 \over 12 }\, \* \z7
          - { 41143 \over 24 }\, \* \z4
          + { 15056 \over 3 }\, \* \z5
          + { 16665 \over 16 }\, \* \z3\nn\\
&&          
          - 829\, \* \zts
          \Big]
       + \caf\, \* \nf\, \* T_R \, \*  \Big[
           { 262835267 \over 62208 }
          - { 337575 \over 256 }\, \* \z6
          - { 224581 \over 96 }\, \* \z7
          + { 2347015 \over 1152 }\, \* \z4\nn\\
&&          
          + { 5841025 \over 2304 }\, \* \z5
          - { 16523531 \over 6912 }\, \* \z3
          - { 887 \over 384 }\, \* \zts
          \Big]
       + \cai \, \*  \Big[
          - { 630869659 \over 497664 }
          + { 2621575 \over 3072 }\, \* \z6\nn\\
&&          
          + { 194793683 \over 98304 }\, \* \z7
          - { 1697837 \over 4608 }\, \* \z4
          - { 124549507 \over 36864 }\, \* \z5
          + { 77047459 \over 110592 }\, \* \z3
          + { 567483 \over 4096 }\, \* \zts
          \Big]\,,
\eea

\bea
(\gamma_3^c)_0 &&=
       - { 3 \over 4 } \ca\,,\qquad 
(\gamma_3^c)_1\;=
        { 5 \over 6 } \ca\, \* \nf\, \* T_R 
       - { 95 \over 48 } \cas \,, 
\eea
\bea
(\gamma_3^c)_2&&=
        { 35 \over 27 } \ca\, \* \nfs\, \* T_R^2 
       + \ca\, \* \cf\, \* \nf\, \* T_R \, \*  \Big[
           { 45 \over 4 }
          - 12\, \* \z3
          \Big]\nn \\
&&
       + \cas\, \* \nf\, \* T_R \, \*  \Big[
           { 97 \over 108 }
          + 9\, \* \z3
          \Big]       + \cat \, \*  \Big[
          - { 15817 \over 1728 }
          - { 9 \over 32 }\, \* \z3
          \Big]\,,
\eea
\bea
(\gamma_3^c)_3&&=
        \dfAAna \, \*  \Big[
           { 69 \over 32 }
          + { 10185 \over 128 }\, \* \z5
          - { 609 \over 8 }\, \* \z3
          \Big]
       + \nf\, \* \dfFAna \, \*  \Big[
          - 60\, \* \z5
          + 48\, \* \z3
          \Big]\nn \\
&&
       + \ca\, \* \nft\, \* T_R^3 \, \*  \Big[
           { 166 \over 81 }
          - { 32 \over 9 }\, \* \z3
          \Big]
       + \ca\, \* \cf\, \* \nfs\, \* T_R^2 \, \*  \Big[
          - { 115 \over 27 }
          - 24\, \* \z4
          + 40\, \* \z3
          \Big] \\
          &&
       + \ca\, \* \cfs\, \* \nf\, \* T_R \, \*  \Big[
          - { 271 \over 12 }
          + 120\, \* \z5
          - 74\, \* \z3
          \Big]
       + \cas\, \* \nfs\, \* T_R^2 \, \*  \Big[
          - { 628 \over 81 }
          + 18\, \* \z4
          - 30\, \* \z3
          \Big]\nn \\
&&
       + \cas\, \* \cf\, \* \nf\, \* T_R \, \*  \Big[
           { 13171 \over 216 }
          + 66\, \* \z4
          - 60\, \* \z5
          - 88\, \* \z3
          \Big]
       + \cat\, \* \nf\, \* T_R \, \*  \Big[
           { 295855 \over 5184 }\nn \\
&&
          - { 801 \over 16 }\, \* \z4
          - 55\, \* \z5
          + { 3019 \over 24 }\, \* \z3
          \Big]
       + \caf \, \*  \Big[
          - { 319561 \over 4608 }
          + { 99 \over 64 }\, \* \z4
          + { 47665 \over 1024 }\, \* \z5
          - { 140743 \over 4608 }\, \* \z3
          \Big]\,,\nn
\eea
\bea
(\gamma_3^c)_4&&=
        \nf\, \* T_R\, \* \dfAAna \, \*  \Big[
          - { 191 \over 24 }
          + { 16975 \over 64 }\, \* \z6
          - { 5285 \over 8 }\, \* \z7
          - { 609 \over 4 }\, \* \z4
          + { 11615 \over 64 }\, \* \z5
          + { 52709 \over 192 }\, \* \z3\nn \\
&&
          + { 2017 \over 32 }\, \* \zts
          \Big]
       + \nfs\, \* T_R\, \* \dfFAna \, \*  \Big[
          - 200\, \* \z6
          + 96\, \* \z4
          + { 1000 \over 3 }\, \* \z5
          - { 416 \over 3 }\, \* \z3
          - 48\, \* \zts
          \Big]\nn \\
&&
       + \cf\, \* \nf\, \* \dfFAna \, \*  \Big[
           { 8 \over 3 }
          + 238\, \* \z7
          - 470\, \* \z5
          + 44\, \* \z3
          + 144\, \* \zts
          \Big]\nn \\
&&
       + \ca\, \* \dfAAna \, \*  \Big[
          - { 7145 \over 2304 }
          - { 186725 \over 256 }\, \* \z6
          - { 6235831 \over 8192 }\, \* \z7
          + { 5643 \over 16 }\, \* \z4
          + { 3520375 \over 1536 }\, \* \z5\nn \\
&&
          - { 4069205 \over 3072 }\, \* \z3
          + { 16887 \over 1024 }\, \* \zts
          \Big]
       + \ca\, \* \nf\, \* \dfFAna \, \*  \Big[
          - { 260 \over 3 }
          + 550\, \* \z6
          + 392\, \* \z7
          - 108\, \* \z4\nn \\
&&
          - { 11545 \over 6 }\, \* \z5
          + { 11395 \over 6 }\, \* \z3
          - { 893 \over 2 }\, \* \zts
          \Big]
       + \ca\, \* \nfs\, \* \dfFFna \, \*  \Big[
           { 428 \over 3 }
          - 48\, \* \z4
          + 240\, \* \z5\nn \\
&&
          - { 436 \over 3 }\, \* \z3
          - 128\, \* \zts
          \Big]
       + \ca\, \* \nff\, \* T_R^4 \, \*  \Big[
           { 260 \over 81 }
          - { 64 \over 9 }\, \* \z4
          + { 320 \over 81 }\, \* \z3
          \Big]\nn \\
&&
       + \ca\, \* \cf\, \* \nft\, \* T_R^3 \, \*  \Big[
          - { 14765 \over 486 }
          + 80\, \* \z4
          - { 128 \over 3 }\, \* \z5
          - { 232 \over 9 }\, \* \z3
          \Big]
       + \ca\, \* \cfs\, \* \nfs\, \* T_R^2 \, \*  \Big[
          - { 53927 \over 216 }\nn\eea
\bea          
\phantom{(\gamma_3^c)_4}
&&
          + 400\, \* \z6
          - 198\, \* \z4
          - { 2000 \over 3 }\, \* \z5
          + { 7588 \over 9 }\, \* \z3
          - 224\, \* \zts
          \Big]
       + \ca\, \* \cft\, \* \nf\, \* T_R \, \*  \Big[
           21
          - 1680\, \* \z7
          \nn\\
  &&
          + 1470\, \* \z5
          + 78\, \* \z3
          \Big]
       + \cas\, \* \nft\, \* T_R^3 \, \*  \Big[
          - { 23837 \over 972 }
          - { 364 \over 9 }\, \* \z4
          + { 928 \over 9 }\, \* \z5
          - { 3820 \over 81 }\, \* \z3
          \Big]\nn \\
&&
       + \cas\, \* \cf\, \* \nfs\, \* T_R^2 \, \*  \Big[
          - { 27380 \over 243 }
          - 200\, \* \z6
          - 476\, \* \z4
          + 568\, \* \z5
          - { 976 \over 9 }\, \* \z3
          + 464\, \* \zts
          \Big]\nn\\
&&
       + \cas\, \* \cfs\, \* \nf\, \* T_R \, \*  \Big[
           { 3119 \over 108 }
          - 1100\, \* \z6
          + 1680\, \* \z7
          + 396\, \* \z4
          + { 2065 \over 3 }\, \* \z5
          - { 27419 \over 36 }\, \* \z3
          - 440\, \* \zts
          \Big]\nn \\
&&
       + \cat\, \* \nfs\, \* T_R^2 \, \*  \Big[
           { 41593 \over 1944 }
          - { 550 \over 3 }\, \* \z6
          + { 5995 \over 12 }\, \* \z4
          + { 157 \over 24 }\, \* \z5
          - { 187019 \over 432 }\, \* \z3
          - { 664 \over 3 }\, \* \zts
          \Big]\nn \\
&&
       + \cat\, \* \cf\, \* \nf\, \* T_R \, \*  \Big[
           { 345017 \over 972 }
          + 550\, \* \z6
          - { 16681 \over 24 }\, \* \z7
          + { 11309 \over 16 }\, \* \z4
          - 1056\, \* \z5
          - { 19001 \over 32 }\, \* \z3\nn \\
&&
          + { 829 \over 2 }\, \* \zts
          \Big]
       + \caf\, \* \nf\, \* T_R \, \*  \Big[
           { 60298109 \over 124416 }
          + { 337575 \over 512 }\, \* \z6
          + { 224581 \over 192 }\, \* \z7
          - { 2123143 \over 2304 }\, \* \z4\nn \\
&&
          - { 8674945 \over 4608 }\, \* \z5
          + { 46881313 \over 41472 }\, \* \z3
          + { 887 \over 768 }\, \* \zts
          \Big]
       + \cai \, \*  \Big[
          - { 47894269 \over 110592 }
          - { 2621575 \over 6144 }\, \* \z6\nn \\
&&
          - { 194793683 \over 196608 }\, \* \z7
          + { 1604909 \over 9216 }\, \* \z4
          + { 128829827 \over 73728 }\, \* \z5
          - { 224465897 \over 663552 }\, \* \z3
          - { 567483 \over 8192 }\, \* \zts
          \Big]\,.
\eea

These results are valid for any gauge group which is a simple Lie group
and are expressed in terms of Casimirs. The notation for these Casimirs is explained in detail in \cite{vanRitbergen:1998pn}.
For fermions transforming in the fundamental representation and the standard normalisation of the SU($N$) generators, 
the Casimirs evaluate to
\bea
\label{colSU(N)}
 &&
\qquad\qquad\qquad
 N_R\;=\;N
\; , \quad 
 N_A\;=\;N^2-1
\; , \quad 
 T_R \:=\: \frac{1}{2} 
\; , \nn \\[1mm] && 
 C_A \:=\: N 
\; , \quad  
 C_F \:=\: \frac{N_A}{2 N} \:=\: \frac{N^2-1}{2 N} 
 \; , \quad 
 \dfAAna \:=\: \frac{N^2(N^2+36)}{24}
\; ,  \\[1mm] && 
\qquad
\dfFAna \:=\: \frac{ N(N^2+6)}{48}
\; , \quad
 \dfFFna \:=\: \frac{N^4-6N^2+18}{96\, N^2} 
\; .\nn
\eea

 We have checked that in the Landau gauge the leading terms in $n_f$ for quark, gluon, and ghost ADs agree with the results in \cite{Gracey:1993ua} while the subleading $n_f$-term of the quark ADs are in agreement with the results in \cite{Ciuchini:1999wy}. We further confirm that in the Landau gauge we have $Z_{1}^{\mathrm{ccg}}=1$ \cite{Taylor:1971ff,Blasi:1990xz}, constituting yet another non-trivial cross check of our calculation. Results of all the ADs and RCs valid for arbitrary gauges can be found in an ancillary file in computer-readable format with the \verb+arxiv+ submission of this paper.

\section{Applications}
\label{sec:Applications}
\subsection{Scheme independent propagators}	
In this section we apply our results to compute the MS-scheme independent gluon, ghost and fermion propagators to N${}^4$LO in perturbative QCD. We thereby extend the N${}^3$LO results which were, for gluon and ghost propagators, presented in \cite{Chetyrkin:2004mf} to N${}^4$LO. Given the \MSb-renormalised propagator $G(\mu,a(\mu))$ with renormalisation group equation,
\beq
\frac{d G(\mu,a)}{d\log\mu^2} =\gamma(a)G(\mu,a)\,,
\eeq
one defines the scale invariant quantity $\hat G(\mu,a)$ such that 
\beq
\hat G(\mu,a) =G(\mu,a)/f(a)\,,\qquad \frac{d\hat G(\mu,a)}{d\log\mu^2} =0\,.
\eeq
It follows that the function $f(a)$ satisfies the differential equation, 
\beq
 \gamma(a)\,f(a)-\beta(a)\,a \frac{d}{d a}\,f(a)=0\,,
\eeq
which is formally solved by
\beq
\label{eq:fscale}
f(a)=\exp\bigg(\int \frac{da}{a} \frac{\gamma(a)}{\beta(a)}\bigg)\,.
\eeq
Up to an overall exponential factor the function $f(a)$ can be expanded perturbatively as
\bea
f(a)&&=a^{-\gamma_0/\beta_0}\Big[ 1+ \bar\gamma_1\,a +\big(\bar \gamma_2-\bar\gamma_1\bar\beta_1+\bar\gamma_1^2\big)\,\frac{a^2}{2}\nn\\
&&+\Big(2 \bar\beta_1^2 \bar\gamma_1-3 \bar\beta_1 \bar\gamma_1^2+\bar\gamma_1^3-2 \bar\beta_1 \bar\gamma_2-2 \bar\beta_2 \bar\gamma_1+3 \bar\gamma_1 \bar\gamma_2+2 \bar\gamma_3\Big)\,\frac{a^3}{3!}\\
&&+\Big(-6 \bar\beta_1^3 \bar\gamma_1+11 \bar\beta_1^2 \bar\gamma_1^2-6 \bar\beta_1 \bar\gamma_1^3+\bar\gamma_1^4+6 \bar\beta_1^2 \bar\gamma_2+12 \bar\beta_1 \bar\beta_2 \bar\gamma_1 -14 \bar\beta_1 \bar\gamma_1 \bar\gamma_2\nn\\
  &&\quad\;\; -8 \bar\beta_2 \bar\gamma_1^2+6 \bar\gamma_1^2 \bar\gamma_2-6 \bar\beta_1 \bar\gamma_3-6 \bar\beta_2 \bar\gamma_2-6 \bar\beta_3 \bar\gamma_1+8 \bar\gamma_1 \bar\gamma_3+3 \bar\gamma_2^2+6 \bar\gamma_4\Big)\,\frac{a^4}{4!}+\order{(a^5)}\Big]\nn
\eea
where $\bar\beta_i=\beta_i/\beta_0$ and $\bar\gamma_i=(\gamma_i-\bar\beta_i\gamma_0)/\beta_0$.

We proceed by defining the scale invariant gluon, ghost and massless quark propagators:
\bea
\hat D^{-1}(q^2) &=& f(a)(1+\Pi(q^2))\,,\qquad \hat D^{ab}_{\mu\nu}=\frac{\delta^{ab}}{-q^2}\bigg[-g_{\mu\nu}+\frac{q_\mu q_\nu}{q^2}\bigg] \hat D(q^2)\,,\nn\\
\hat \Delta^{-1}(q^2) &=& f(a)(1+\widetilde \Pi(q^2))\,,\qquad \hat \Delta^{ab}=\frac{\delta^{ab}}{-q^2}\hat \Delta(q^2)\,,\\ 
\hat S^{-1}(q^2) &=& f(a)(1+\Sigma(q^2))\,,\qquad\hat S^{ij}=\frac{\delta^{ij}}{-q^2} \hat S(q^2)\,.\nn
\eea
where the scalar functions $\hat D(q^2)$, $\hat \Delta(q^2)$ and $\hat S(q^2)$ are related to the 1PI self energies.
We provide the corresponding numeric values for the SU($3$) gauge group by replacing $\gamma=\gamma_3,\gamma_3^c,\gamma_2$ respectively in eq. (\ref{eq:fscale}) and using the four-loop self energy published in \cite{Ruijl:2017eht}. The gluon propagator at different number of quark flavours is
\bea
 a^{\frac{13}{22}}\hat D^{-1}(q^2)|_{n_F=0} &=&1-2.15952\,a_s-11.6762\,a_s^2-88.8523\,a_s^3-987.996\,a_s^4,\nn\\
 a^{\frac{1}{2}}\hat D^{-1}(q^2)|_{n_F=3} &=&1-1.29514\,a_s-5.31783\,a_s^2-34.0328\,a_s^3-287.942\,a_s^4,\nn\\
a^{\frac{19}{46}}\hat D^{-1}(q^2)|_{n_F=5} &=& 1-0.617944\,a_s-1.13804\,a_s^2-7.32128\,a_s^3-25.9219\,a_s^4,\nn\\
a^{\frac{5}{14}}\hat D^{-1}(q^2)|_{n_F=6} &=& 1-0.214498\,a_s+1.03846\,a_s^2+3.44460\,a_s^3+51.6541\,a_s^4,
\eea
where $a_s=\frac{g^2}{4\pi^2}=4a$ and we set for convenience $\mu^2=-q^2$. Likewise, we get the ghost propagator
\bea
 a^{\frac{9}{44}}\hat \Delta^{-1}(q^2)|_{n_F=0} &=&1-0.680656\,a_s-4.52931\,a_s^2-38.7849\,a_s^3-466.453\,a_s^4,\nn\\
 a^{\frac{1}{4}}\hat \Delta^{-1}(q^2)|_{n_F=3} &=&1-0.696181\,a_s-3.28694\,a_s^2-21.2361\,a_s^3-195.589\,a_s^4,\nn\\
a^{\frac{27}{92}}\hat \Delta^{-1}(q^2)|_{n_F=5} &=& 1-0.757000\,a_s-2.41477\,a_s^2-10.7375\,a_s^3-70.6130\,a_s^4,\nn\\
a^{\frac{9}{28}}\hat \Delta^{-1}(q^2)|_{n_F=6} &=& 1-0.819834\,a_s-1.97443\,a_s^2-5.81361\,a_s^3-23.5339\,a_s^4,
\eea
and the fermion propagator
\bea
\hat S^{-1}(q^2)|_{n_F=0} &=&1+0.507576\,a_s+2.63292\,a_s^2+27.0067\,a_s^3+330.042\,a_s^4,\nn\\
 \hat S^{-1}(q^2)|_{n_F=3} &=&1+0.509259\,a_s+2.05204\,a_s^2+14.8910\,a_s^3+138.286\,a_s^4,\nn\\
\hat S^{-1}(q^2)|_{n_F=5} &=& 1+0.510870\,a_s+1.65428\,a_s^2+7.77199\,a_s^3+51.1008\,a_s^4,\nn\\
\hat S^{-1}(q^2)|_{n_F=6} &=& 1+0.511905\,a_s+1.45060\,a_s^2+4.49450\,a_s^3+18.7354\,a_s^4.
\eea
The perturbative convergence of all three scaleless propagators appears within reason for low values of $n_f$ and shows improvement as $n_f$ is increased. This feature will be lost as $n_f$ becomes closer to the value $n_f\sim 16.5$ where a singularity in $\frac{1}{\beta_0}$ forces a consequent loss of perturbativity in that region. 

\subsection{Anomalous Dimension of the operator $A^2$}
In Landau gauge the AD of the composite operator $A^2$ is related to the beta function and the AD of the gluon wave function \cite{Gracey:2002yt,Dudal:2003np,Dudal:2002pq,Dudal:2004ch}:
\beq
-2\gamma_A\big|_{\xi_L=0}=\beta-\gamma_3\,.
\eeq
This result was given at the 4-loop level in \cite{Chetyrkin:2004mf}. 
Here, specialising our result to QCD with gauge group SU($3$), we extend it to 5 loop level:  
\bea
\gamma_A\big|_{\xi_L=0}&=&
a\, \* \Big(-{ 2 \over 3 }\, \* \nf +{ 35 \over 4 }\Big)
+a^2\, \*\Big({ 1347 \over 16 }-{ 137 \over 12 }\, \* \nf\Big)\,\nn\\
&
+&a^3\, \*\Big(
{ 75607 \over 64 }-{ 243 \over 32 }\, \* \z3
+\Big[{ 33 \over 2 }\, \* \z3-{ 18221 \over 72 }\Big]\, \* \nf
+{ 755 \over 108 }\, \* \nfs
\Big) 
\nn\\
&
+&a^4\, \*\Big(
{ 40905 \over 8 }\, \* \z5-{ 99639 \over 512 }\, \* \z3+{ 8019 \over 64 }\, \* \z4+{ 29764511 \over 1536 }\nn\\
&&
\quad +\Big[-{ 3355 \over 4 }\, \* \z5-{ 8955 \over 32 }\, \* \z4+{ 335585 \over 432 }\, \* \z3-{ 57858155 \over 10368 }\Big]\, \* \nf\\
&&
\quad+\Big[{ 8489 \over 162 }\, \* \z3+{ 33 \over 2 }\, \* \z4+{ 46549 \over 162 }\Big]\, \* \nfs
+\Big[-{ 4 \over 3 }\, \* \z3+{ 6613 \over 2916 }\Big]\, \* \nft
\Big)\, \*  \nn\\
&+&
a^5\, \* \Big(
{ 1656617009 \over 4096 }+{ 662250297 \over 4096 }\, \* \z3+{ 16452855 \over 1024 }\, \* \z4+{ 2068873299 \over 8192 }\, \* \z5\nn\\
&&
\quad
-{ 2249775 \over 16 }\, \* \z6-{ 9151899939 \over 32768 }\, \* \z7-{ 65529567 \over 4096 }\, \* \zts
+\Big[
{ 440809 \over 128 }\, \* \zts-{ 66794279 \over 1728 }\, \* \z5\nn\\
&&
\quad
+{ 3662827 \over 96 }\, \* \z7-{ 26669221 \over 1536 }\, \* \z4
+{ 63175 \over 2 }\, \* \z6
-{ 929208143 \over 41472 }\, \* \z3-{ 36593462075 \over 248832 }
\Big]\, \* \nf\nn\\
&&\quad
+\Big[
-{ 16775 \over 12 }\, \* \z6-{ 2659 \over 6 }\, \* \zts+{ 33243821 \over 5184 }\, \* \z3+{ 78931 \over 144 }\, \* \z4-{ 9486899 \over 2592 }\, \* \z5\nn\eea
\bea          
\phantom{\gamma_A\big|_{\xi_L=0}}
&&\quad+{ 99923027 \over 7776 }
\Big]\, \* \nfs
+\Big[{ 1312 \over 9 }\, \* \z5+{ 2939 \over 54 }\, \* \z4-{ 129493 \over 486 }\, \* \z3-{ 3520195 \over 23328 }
\Big]\, \* \nft\qquad\qquad\nn\\
&&\quad
+\Big[
-{ 4 \over 3 }\, \* \z4-{ 92 \over 81 }\, \* \z3+{ 740 \over 729 }
\Big]\, \* \nff
\Big)+\order(a^6)\nn\\
\eea

\section{Summary}
\label{Conclusions}

In this work we complete the renormalisation of non-abelian gauge theory 
with $n_f$ fermions in an arbitrary representation to five-loop order for a generic covariant gauge 
and an arbitrary simple gauge group. In particular we have extended the results of
\cite{Baikov:2014qja, Baikov:2016tgj,Luthe:2016ima,Luthe:2016xec,Herzog:2017bjx,Luthe:2017ttc,Baikov:2017ujl}, 
which were given in Feynman gauge only, by obtaining the complete dependence on the gauge-fixing parameter. 
All results are included in the \verb+arxiv+ submission of this paper as ancillary file in computer-readable format. 
Furthermore we applied our results to compute the five-loop Landau gauge scale-independent gluon, ghost and 
fermion propagators as well as the Landau gauge AD of the composite operator $A^2$. 

\textbf{Remark:} on the day on which this article was submitted, ref. \cite{Luthe:2017ttg} appeared where all the anomalous dimensions are computed in expansion around the Feynman gauge $\xi=0$, including the linear terms. Compared to this work, ref. \cite{Luthe:2017ttg} follows an entirely different approach, involving the calculation of completely massive tadpoles at five loops. We verified that our results are in agreement with those reported in ref. \cite{Luthe:2017ttg}, which is an important independent check on the ADs.


\textbf{Remark 2:} all scheme independent propagators considered in subsection 4.1 depend on only one even zeta, namely $\zeta_4$, which occurs {\em only} at order $a^4$ . All terms which contain $\zeta_6$ in the 4-loop propagators  and in the corresponding 5-loop anomalous dimensions neatly cancel each other in the scheme independent  combinations  considered in subsection 4.1. Analytic results for these  scheme independent propagators in Landau gauge have been added in an ancillary file in computer-readable format with the \verb+arxiv+ submission of this paper. Even more, after transition to  the so-called C-scheme \cite{Boito:2016pwf} the results are fully free from any even zetas. The  same is valid if one  constructs scheme invariant versions of the vertex functions computed in \cite{Ruijl:2017eht} (again taken in  the Landau gauge). The phenomenon was first discovered in \cite{Jamin:2017mul} on examples of the scalar and gluon correlators that enter the  hadronic decays  of the Higgs  boson. Many  more  examples of  absence of $\pi^2$ terms in so-called physical anomalous dimensions in DIS have been very recently discussed in \cite{Davies:2017hyl}.

\textbf{Note added:} by the time this paper was published, we completed the calculation of the quark-gluon vertex to five loops with all the powers of $\xi$. This result was used to re-compute the renormalisation constant $Z_1^{\mathrm{\psi\psi g}}$, which was originally derived from $Z_g$, $Z_3$ and $Z_2$ by applying the Slavnov-Taylor identities in eq.(\ref{WI:qqg}). The results of the two procedures agree with each other, thus providing another check on our calculation.

\section*{Acknowledgements}

We would like to thank T. Ueda and B. Ruijl for help with {\sc FORM} and the {\sc Forcer} package. 
We thank A. Vogt for comparing with us the result of an independent computation of the linear gauge dependence of $\gamma_2$ 
which constitutes another cross check for our result. We are greatful for the help of J. Gracey for providing cross-checks for parts 
of the results presented here. This work is supported by the ERC Advanced Grant no. 320651, ``HEPGAME''. The work by K.~G.~Chetyrkin was supported by the Deutsche Forschungsgemeinschaft through CH1479/1-1 and by the German Federal Ministry for Education and Research BMBF through Grant No.~05H15GUCC1.

\glsaddall
\printglossaries

\bibliographystyle{JHEP}
\bibliography{refs}
 
\end{document}